\documentclass[twocolumn,showpacs,preprintnumbers,amsmath,amssymb]{revtex4}
\usepackage{graphicx}% Include figure files
\usepackage{dcolumn}% Align table columns on decimal point
\usepackage{bm}% bold math
\usepackage{amsthm}
\begin{document}
\title{\bf{Simultaneous Dense Coding}}
\author{Haozhen Situ$^{1}$}\author{Daowen Qiu$^{1,2}$}\email{issqdw@mail.sysu.edu.cn}%

 \affiliation{%
 $^{1}$Department of
Computer Science, Zhongshan University, Guangzhou 510275,
 People's Republic of China\\
 $^{2}$SQIG--Instituto de Telecomunica\c{c}\~{o}es, IST,
TULisbon,  Av. Rovisco Pais 1049-001, Lisbon, Portugal
}%

\date{\today}

 \begin{abstract}
We present a dense coding scheme between one sender and two
receivers, which guarantees that the receivers simultaneously obtain
their respective messages. In our scheme, the quantum entanglement
channel is first locked by the sender so that the receivers cannot
learn their messages unless they collaborate to perform the
unlocking operation. We also show that the quantum Fourier transform
can act as the locking operator both in simultaneous dense coding
and teleportation.
 \end{abstract}
  \maketitle

\section{\label{Introduction} Introduction}
Quantum entanglement \cite{HHHH09} is the key resource of quantum
information theory \cite{NC00,ABH01}, especially in quantum
communication \cite{GT07}. Sharing an entangled quantum state
between a sender and a receiver, makes it possible to perform
quantum teleportation \cite{BBCJPW93} and quantum dense coding
\cite{BW92}. Quantum teleportation is the process of transmitting an
unknown quantum state by using shared entanglement and sending
classical information; quantum dense coding is the process of
transmitting 2 bits of classical information by sending part of an
entangled state. Teleportation and dense coding are closely related
\cite{W01,HLG00} and have been extensively studied in various ways.
For example, teleportation and dense coding that use non-maximally
entangled quantum channel have been examined
\cite{LLG00,AP02,PA04,GR06,BE95,HJSWW96,HLG00,B01,MOR05,PPA05};
multipartite entangled states have also been considered as the
quantum channel
\cite{MW00,GT00,GTRZ03,JPOK03,BVK98,LLTL02,BLSSDM06,AP06,LQ07};
another generalization is to perform these two communication tasks
under the control of a third party, so called controlled
teleportation and dense coding
\cite{KB98,DLLZW05,MXA07,LQL09,HLG01,LO09}.

Recently, a simultaneous quantum state teleportation scheme was
proposed by Wang et al \cite{WYGL08}, the aim of which is for all
the receivers to simultaneously obtain their respective quantum
states from Alice (the sender). In their scheme, Alice first performs
a unitary transform to lock the entanglement channel, and therefore the
receivers cannot restore their quantum states separately before
performing an unlocking operation together.  A natural question is
that whether this idea of locking the entanglement channel adapts
for dense coding? The main purpose of this paper is to show that
such a locking operator for dense coding really exists. As a result,
we propose three simultaneous dense coding protocols which guarantee
that the receivers simultaneously obtain their respective messages.

The remainder of the paper is organized as follows. In Sec.
\ref{Protocols}, we introduce three simultaneous dense coding
protocols using different entanglement channels. In Sec.
\ref{Teleportation}, we show that the quantum Fourier transform can
alternatively be used as the locking operator in simultaneous
teleportation. A brief conclusion follows in Sec. \ref{Conclusion}.

\section{\label{Protocols} Protocols for Simultaneous Dense Coding}
Suppose that Alice is the sender, Bob and Charlie are the receivers.
Alice intends to send two bits $(b_1,b_2)$ to Bob and another two
bits $(c_1,c_2)$ to Charlie under the condition that Bob and Charlie
must collaborate to simultaneously find out what she sends.

In the following three subsections, we propose three protocols using
Bell state, GHZ state and W state as the entanglement channels
respectively. The idea of these protocols is to perform the quantum
Fourier transform on Alice's qubits before sending them to Bob and
Charlie. After receiving Alice's qubits, Bob and Charlie's local
states are independent of $(b_1,b_2)$ and $(c_1,c_2)$ so that they
know nothing about the encoded bits. Only after performing the
inverse quantum Fourier transform together, they can obtain
$(b_1,b_2)$ and $(c_1,c_2)$ respectively.

\subsection{Protocol 1: Using Bell State}
Initially, Alice, Bob and Charlie share two
Einstein-Podolsky-Rosen(EPR) pairs \cite{EPR35}
$\frac{1}{\sqrt{2}}(|00\rangle +|11\rangle)_{A_1B}$ and
$\frac{1}{\sqrt{2}}(|00\rangle +|11\rangle)_{A_2C}$, where qubits
$A_1A_2$ belong to Alice, qubits $B$ and $C$ belong to Bob and
Charlie respectively. The initial quantum state of the composite
system is
\begin{align}
|\psi(0)\rangle = \frac{1}{\sqrt{2}}(|00\rangle
+|11\rangle)_{A_1B}\otimes \frac{1}{\sqrt{2}}(|00\rangle
+|11\rangle)_{A_2C}.
\end{align}

The protocol consists of four steps.

(1) Alice performs unitary transforms $U(b_1b_2)$ on qubits $A_1$
and $U(c_1c_2)$ on $A_2$ to encode her bits, like the original dense
coding scheme \cite{BW92}. After that, the state of the composite
system becomes
\begin{align}|\psi(1)\rangle = U_{A_1}(b_1b_2)\otimes
U_{A_2}(c_1c_2)|\psi(0)\rangle \nonumber\\
=
|\phi(b_1b_2)\rangle_{A_1B} \otimes |\phi(c_1c_2)\rangle_{A_2C},
\end{align} where
\begin{align}
& U(00) = I =
\begin{bmatrix} 1 & 0 \\ 0 & 1 \end{bmatrix}, U(01)
 = \sigma_z = \begin{bmatrix} 1 & 0 \\ 0 & -1 \end{bmatrix},\nonumber\\
& U(10) = \sigma_x = \begin{bmatrix} 0 & 1 \\ 1 & 0 \end{bmatrix},
U(11) = \sigma_z\sigma_x = \begin{bmatrix} 0 & 1 \\ -1 & 0
 \end{bmatrix}, \nonumber\\
& |\phi (xy)\rangle=
\frac{1}{\sqrt{2}}(|0x\rangle+(-1)^{y}|1\overline{x}\rangle).
\end{align}

(2) Alice performs the quantum Fourier transform
\begin{align}
QFT = \frac{1}{2} \begin{bmatrix} 1 & 1 & 1 & 1\\
1 & i & -1 & -i \\ 1 & -1 & 1 & -1 \\ 1 & -i & -1 & i
\end{bmatrix}\end{align} on qubits $A_1A_2$ to lock the entanglement channel, and then sends $A_1$ to Bob and $A_2$ to
Charlie. The state of the composite system becomes
\begin{align}
|\psi(2)\rangle = QFT_{A_1A_2}[|\phi(b_1b_2)\rangle_{A_1B} \otimes
|\phi(c_1c_2)\rangle_{A_2C}].
\end{align}

(3) Bob and Charlie collaborate to perform $QFT^\dagger$ on qubits
$A_1A_2$. The state of the composite system becomes
\begin{align}
|\psi(3)\rangle = & QFT_{A_1A_2}^\dagger
QFT_{A_1A_2}[|\phi(b_1b_2)\rangle_{A_1B} \otimes
|\phi(c_1c_2)\rangle_{A_2C}] \nonumber\\
= & |\phi(b_1b_2)\rangle_{A_1B} \otimes |\phi(c_1c_2)\rangle_{A_2C}.
\end{align}

(4) Bob and Charlie perform the Bell State Measurement on qubits
$A_1B$ and $A_2C$ respectively to obtain $(b_1,b_2)$ and
$(c_1,c_2)$, like the original dense coding scheme \cite{BW92}.

The following theorem demonstrates that neither Bob nor Charlie
alone can distinguish his two-qubit quantum state (i.e. $\rho_{A_1B},
\rho_ {A_2C}$) before step 3. Therefore, they cannot learn the
encoded bits from their quantum states unless they collaborate.

{\it Theorem 1.} For each $b_1,b_2,c_1,c_2\in\{0,1\}, \
\rho_{A_1B}=\rho_{A_2C}=I/4$, where $\rho_{A_1B}$ and $\rho_{A_2C}$
are the reduced density matrices in subsystems $A_1B$ and $A_2C$
after step 2 (but before step 3).

{\it Proof.} After step 1, the quantum states of qubits $A_1B$ and
$A_2C$ are $|\phi(b_1b_2)\rangle$ and $|\phi(c_1c_2)\rangle$
respectively. The state of the composite system after step 1 can be
written as
\begin{align}
|\psi(1)\rangle =|\phi(b_1b_2)\rangle_{A_1B}\otimes
|\phi(c_1c_2)\rangle_{A_2C}.
\end{align}
After step 2, the state of the composite system becomes
\begin{align}
|\psi(2)\rangle = & QFT_{A_1A_2}
[\frac{1}{\sqrt{2}}(|0b_1\rangle+(-1)^{b_2}|1\overline{b_1}\rangle)_{A_1B}\nonumber\\
& \otimes \frac{1}{\sqrt{2}}
(|0c_1\rangle+(-1)^{c_2}|1\overline{c_1}\rangle)_{A_2C}]\nonumber\\
= &
\frac{1}{2}QFT_{A_1A_2}(|00b_1c_1\rangle+(-1)^{c_2}|01b_1\overline{c_1}\rangle\nonumber\\
&
+(-1)^{b_2}|10\overline{b_1}c_1\rangle+(-1)^{b_2+c_2}|11\overline{b_1c_1}\rangle)_{A_1A_2BC}
\nonumber\\
= &
\frac{1}{4}[(|00\rangle+|01\rangle+|10\rangle+|11\rangle)|b_1c_1\rangle+(-1)^{c_2}(|00\rangle\nonumber\\
&
+i|01\rangle-|10\rangle-i|11\rangle)|b_1\overline{c_1}\rangle+(-1)^{b_2}(|00\rangle\nonumber\\
&-|01\rangle+|10\rangle-|11\rangle)|\overline{b_1}c_1\rangle+(-1)^{b_2+c_2}(|00\rangle\nonumber\\
&-i|01\rangle
-|10\rangle+i|11\rangle)|\overline{b_1c_1}\rangle]_{A_1A_2BC}.
\end{align} The reduced density matrix in subsystem $A_1B$ is
\begin{align}
\rho_{A_1B} = &  _{A_2C}\langle
0c_1|\psi(2)\rangle\langle\psi(2)|0c_1\rangle_{A_2C} +
_{A_2C}\langle
0\overline{c_1}|\psi(2)\rangle\nonumber\\&\langle\psi(2)|0\overline{c_1}\rangle_{A_2C}
+ _{A_2C}\langle
1c_1|\psi(2)\rangle\langle\psi(2)|1c_1\rangle_{A_2C} \nonumber\\&
+
_{A_2C}\langle
1\overline{c_1}|\psi(2)\rangle\langle\psi(2)|1\overline{c_1}\rangle_{A_2C}\nonumber\\
= & \frac{1}{4}(|0b_1\rangle\langle
0b_1|+|0\overline{b_1}\rangle\langle
0\overline{b_1}|+|1b_1\rangle\langle
1b_1|+|1\overline{b_1}\rangle\langle 1\overline{b_1}|)\nonumber\\
= & I/4.
\end{align}The reduced density matrix in subsystem $A_2C$ is
\begin{align}
\rho_{A_2C} = &  _{A_1B}\langle
0b_1|\psi(2)\rangle\langle\psi(2)|0b_1\rangle_{A_1B} +
_{A_1B}\langle
0\overline{b_1}|\psi(2)\rangle\nonumber\\&\langle\psi(2)|0\overline{b_1}\rangle_{A_1B}+
_{A_1B}\langle
1b_1|\psi(2)\rangle\langle\psi(2)|1b_1\rangle_{A_1B}\nonumber\\& +
_{A_1B}\langle
1\overline{b_1}|\psi(2)\rangle\langle\psi(2)|1\overline{b_1}\rangle_{A_1B}\nonumber\\
= & \frac{1}{4}(|0c_1\rangle\langle
0c_1|+|0\overline{c_1}\rangle\langle
0\overline{c_1}|+|1c_1\rangle\langle
1c_1|+|1\overline{c_1}\rangle\langle 1\overline{c_1}|)\nonumber\\
= & I/4.
\end{align}\qed

\subsection{Protocol 2: Using GHZ State}
Initially, Alice, Bob and Charlie share two
Greenberger-Horne-Zeilinger (GHZ) states \cite{GHZ89}
$\frac{1}{\sqrt{2}}(|000\rangle +|111\rangle)_{A_1B_1B_2}$ and
$\frac{1}{\sqrt{2}}(|000\rangle +|111\rangle)_{A_2C_1C_2}$, where
qubits $A_1A_2$ belong to Alice, qubits $B_1B_2$ and $C_1C_2$ belong
to Bob and Charlie, respectively. The initial quantum state of the
composite system is
\begin{align} |\psi(0)\rangle = &
\frac{1}{{\sqrt{2}}}(|000\rangle
+|111\rangle)_{A_1B_1B_2}\nonumber\\&\otimes
\frac{1}{{\sqrt{2}}}(|000\rangle +|111\rangle)_{A_2C_1C_2}.
\end{align}

The protocol consists of four steps.

(1) Alice performs unitary transforms $U(b_1b_2)$ on qubits $A_1$
and $U(c_1c_2)$ on $A_2$ to encode her bits. After that, the state
of the composite system becomes
\begin{align}|\psi(1)\rangle = & U_{A_1}(b_1b_2)\otimes
U_{A_2}(c_1c_2)|\psi(0)\rangle \nonumber\\= &
|GHZ(b_1b_2)\rangle_{A_1B_1B_2} \otimes
|GHZ(c_1c_2)\rangle_{A_2C_1C_2}, \end{align} where
\begin{align}
|GHZ(xy)\rangle=
\frac{1}{\sqrt{2}}(|0xx\rangle+(-1)^{y}|1\overline{xx}\rangle).
\end{align}

(2) Alice performs the quantum Fourier transform on qubits $A_1A_2$
to lock the entanglement channel, and then sends $A_1$ to Bob and $A_2$
to Charlie. The state of the composite system becomes
\begin{align}
|\psi(2)\rangle = &
QFT_{A_1A_2}[|GHZ(b_1b_2)\rangle_{A_1B_1B_2}\nonumber\\ & \otimes
|GHZ(c_1c_2)\rangle_{A_2C_1C_2}].
\end{align}

(3) Bob and Charlie collaborate to perform $QFT^\dagger$ on qubits
$A_1A_2$. The state of the composite system becomes
\begin{align}|\psi(3)\rangle = & QFT_{A_1A_2}^\dagger
QFT_{A_1A_2}[|GHZ(b_1b_2)\rangle_{A_1B_1B_2} \nonumber\\& \otimes
|GHZ(c_1c_2)\rangle_{A_2C_1C_2}] \nonumber\\
= & |GHZ(b_1b_2)\rangle_{A_1B_1B_2} \otimes
|GHZ(c_1c_2)\rangle_{A_2C_1C_2}. \end{align}

(4) Bob and Charlie make the von Neumann measurement using the
orthogonal states $\{|GHZ(xy)\rangle\}_{xy}$ on qubits $A_1B_1B_2$
and $A_2C_1C_2$ respectively to obtain $(b_1,b_2)$ and $(c_1,c_2)$.

The following theorem demonstrates that neither Bob nor Charlie
alone can distinguish his three-qubit quantum state (i.e.
$\rho_{A_1B_1B_2}, \rho_ {A_2C_1C_2}$) before step 3. Therefore,
they cannot learn the encoded bits from their quantum states unless
they collaborate.

{\it Theorem 2.} $\rho_{A_1B_1B_2}$ and $\rho_{A_2C_1C_2}$ are
independent of $b_1,b_2,c_1,c_2$, where $\rho_{A_1B_1B_2}$ and
$\rho_{A_2C_1C_2}$ are the reduced density matrices in subsystems
$A_1B_1B_2$ and $A_2C_1C_2$ after step 2 (but before step 3), respectively.

{\it Proof.} After step 1, the quantum states of qubits $A_1B_1B_2$
and $A_2C_1C_2$ are $|GHZ(b_1b_2)\rangle$ and $|GHZ(c_1c_2)\rangle$,
respectively. The state of the composite system after step 1 can be
written as
\begin{align}
|\psi(1)\rangle = |GHZ(b_1b_2)\rangle_{A_1B_1B_2}\otimes
|GHZ(c_1c_2)\rangle_{A_2C_1C_2}.
\end{align}
After step 2, the state of the composite system becomes
\begin{align}
|\psi(2)\rangle = &
QFT_{A_1A_2}[\frac{1}{\sqrt{2}}(|0b_1b_1\rangle+(-1)^{b_2}|1\overline{b_1b_1}\rangle)_{A_1B_1B_2}\nonumber\\&\otimes
\frac{1}{\sqrt{2}}
(|0c_1c_1\rangle+(-1)^{c_2}|1\overline{c_1c_1}\rangle)_{A_2C_1C_2}]\nonumber\\
= & \frac{1}{2}QFT_{A_1A_2}(|00\rangle\otimes
|b_1b_1c_1c_1\rangle+(-1)^{c_2}|01\rangle \nonumber\\ & \otimes
|b_1b_1\overline{c_1c_1}\rangle +(-1)^{b_2}|10\rangle\otimes
|\overline{b_1b_1}c_1c_1\rangle \nonumber\\
& +(-1)^{b_2+c_2}|11\rangle\otimes
|\overline{b_1b_1c_1c_1}\rangle)_{A_1A_2B_1B_2C_1C_2}
\nonumber\\
= &
\frac{1}{4}[(|00\rangle+|01\rangle+|10\rangle+|11\rangle)\otimes|b_1b_1c_1c_1\rangle
\nonumber\\ & +(-1)^{c_2}
(|00\rangle+i|01\rangle-|10\rangle-i|11\rangle)\otimes|b_1b_1\overline{c_1c_1}\rangle
\nonumber\\ & +(-1)^{b_2}
(|00\rangle-|01\rangle+|10\rangle-|11\rangle)\otimes|\overline{b_1b_1}c_1c_1\rangle
\nonumber\\ &
+(-1)^{b_2+c_2}(|00\rangle-i|01\rangle-|10\rangle+i|11\rangle)\nonumber\\
& \otimes|\overline{b_1b_1c_1c_1}\rangle]_{A_1A_2B_1B_2C_1C_2}.
\end{align} The reduced density matrix in subsystem $A_1B_1B_2$ is
\begin{align}
\rho_{A_1B_1B_2} = &  _{A_2C_1C_2}\langle
0c_1c_1|\psi(2)\rangle\langle\psi(2)|0c_1c_1\rangle_{A_2C_1C_2}
\nonumber\\& + _{A_2C_1C_2}\langle
0\overline{c_1c_1}|\psi(2)\rangle\langle\psi(2)|0\overline{c_1c_1}\rangle_{A_2C_1C_2}\nonumber\\
& + _{A_2C_1C_2}\langle
1c_1c_1|\psi(2)\rangle\langle\psi(2)|1c_1c_1\rangle_{A_2C_1C_2}
\nonumber\\&+ _{A_2C_1C_2}\langle
1\overline{c_1c_1}|\psi(2)\rangle\langle\psi(2)|1\overline{c_1c_1}\rangle_{A_2C_1C_2}\nonumber\\
= & \frac{1}{4}(|0b_1b_1\rangle\langle
0b_1b_1|+|0\overline{b_1b_1}\rangle\langle 0\overline{b_1b_1}|
\nonumber\\& +|1b_1b_1\rangle\langle
1b_1b_1|+|1\overline{b_1b_1}\rangle\langle 1\overline{b_1b_1}|)\nonumber\\
= & \frac{1}{4}(|000\rangle\langle 000|+|011\rangle\langle
011|+|100\rangle\langle 100|\nonumber\\&+|111\rangle\langle 111|).
\end{align}The reduced density matrix in subsystem $A_2C_1C_2$ is
\begin{align}
\rho_{A_2C_1C_2} = &  _{A_1B_1B_2}\langle
0b_1b_1|\psi(2)\rangle\langle\psi(2)|0b_1b_1\rangle_{A_1B_1B_2}
\nonumber\\&
+ _{A_1B_1B_2}\langle
0\overline{b_1b_1}|\psi(2)\rangle\langle\psi(2)|0\overline{b_1b_1}\rangle_{A_1B_1B_2}\nonumber\\
& + _{A_1B_1B_2}\langle
1b_1b_1|\psi(2)\rangle\langle\psi(2)|1b_1b_1\rangle_{A_1B_1B_2}
\nonumber\\&
+ _{A_1B_1B_2}\langle
1\overline{b_1b_1}|\psi(2)\rangle\langle\psi(2)|1\overline{b_1b_1}\rangle_{A_1B_1B_2}\nonumber\\
= & \frac{1}{4}(|0c_1c_1\rangle\langle
0c_1c_1|+|0\overline{c_1c_1}\rangle\langle
0\overline{c_1c_1}|\nonumber\\&
+|1c_1c_1\rangle\langle
1c_1c_1|+|1\overline{c_1c_1}\rangle\langle 1\overline{c_1c_1}|)\nonumber\\
= & \frac{1}{4}(|000\rangle\langle 000|+|011\rangle\langle
011|+|100\rangle\langle 100|\nonumber\\&
+|111\rangle\langle 111|).
\end{align}\qed

\subsection{Protocol 3: Using W State}
Initially, Alice, Bob and Charlie share two W states
\cite{DVC00,AP06} $\frac{1}{2}(|010\rangle
+|001\rangle+\sqrt{2}|100\rangle)_{A_1B_1B_2}$ and
$\frac{1}{2}(|010\rangle
+|001\rangle+\sqrt{2}|100\rangle)_{A_2C_1C_2}$, where qubits
$A_1A_2$ belong to Alice, qubits $B_1B_2$ and $C_1C_2$ belong to Bob
and Charlie, respectively. The initial quantum state of the composite
system is
\begin{align}
|\psi(0)\rangle = & \frac{1}{2}(|010\rangle
+|001\rangle+\sqrt{2}|100\rangle)_{A_1B_1B_2}\nonumber\\& \otimes
\frac{1}{2}(|010\rangle
+|001\rangle+\sqrt{2}|100\rangle)_{A_2C_1C_2}.
\end{align}

The protocol consists of four steps.

(1) Alice performs unitary transforms $U(b_1b_2)$ on qubits $A_1$
and $U(c_1c_2)$ on $A_2$ to encode her bits. After that, the state
of the composite system becomes
\begin{align}|\psi(1)\rangle = & U_{A_1}(b_1b_2)\otimes
U_{A_2}(c_1c_2)|\psi(0)\rangle \nonumber\\ = &
|W(b_1b_2)\rangle_{A_1B_1B_2} \otimes |W(c_1c_2)\rangle_{A_2C_1C_2},
\end{align} where
\begin{align}
|W(xy)\rangle=
\frac{1}{2}(|x10\rangle+|x01\rangle+(-1)^{y}\sqrt{2}|\overline{x}00\rangle).
\end{align}

(2) Alice performs the quantum Fourier transform on qubits $A_1A_2$
to lock the entanglement channel, and then sends $A_1$ to Bob and $A_2$
to Charlie. The state of the composite system becomes
\begin{align}
|\psi(2)\rangle = & QFT_{A_1A_2}[|W(b_1b_2)\rangle_{A_1B_1B_2}
\nonumber\\&\otimes |W(c_1c_2)\rangle_{A_2C_1C_2}].
\end{align}

(3) Bob and Charlie collaborate to perform $QFT^\dagger$ on qubits
$A_1A_2$. The state of the composite system becomes
\begin{align}|\psi(3)\rangle = & QFT_{A_1A_2}^\dagger
QFT_{A_1A_2}[|W(b_1b_2)\rangle_{A_1B_1B_2} \nonumber\\& \otimes
|W(c_1c_2)\rangle_{A_2C_1C_2}] \nonumber\\
= & |W(b_1b_2)\rangle_{A_1B_1B_2} \otimes
|W(c_1c_2)\rangle_{A_2C_1C_2}. \end{align}

(4) Bob and Charlie make the von Neumann measurement using the
orthogonal states $\{|W(xy)\rangle\}_{xy}$ on qubits $A_1B_1B_2$ and
$A_2C_1C_2$ respectively to obtain $(b_1,b_2)$ and $(c_1,c_2)$.

The following theorem demonstrates that neither Bob nor Charlie
alone can distinguish his three-qubit quantum state (i.e.
$\rho_{A_1B_1B_2}, \rho_ {A_2C_1C_2}$) before step 3. Therefore,
they cannot learn the encoded bits from their quantum states unless
they collaborate.

{\it Theorem 3.} $\rho_{A_1B_1B_2}$ and $\rho_{A_2C_1C_2}$ are
independent of $b_1,b_2,c_1,c_2$, where $\rho_{A_1B_1B_2}$ and
$\rho_{A_2C_1C_2}$ are the reduced density matrices in subsystems
$A_1B_1B_2$ and $A_2C_1C_2$ after step 2 (but before step 3), respectively.

{\it Proof.} After step 1, the quantum states of qubits $A_1B_1B_2$
and $A_2C_1C_2$ are $|W(b_1b_2)\rangle$ and $|W(c_1c_2)\rangle$
respectively. The state of the composite system after step 1 can be
written as
\begin{align}
|\psi(1)\rangle = |W(b_1b_2)\rangle_{A_1B_1B_2}\otimes
|W(c_1c_2)\rangle_{A_2C_1C_2}.
\end{align}
After step 2, the state of the composite system becomes
\begin{align}
|\psi(2)\rangle = &
QFT_{A_1A_2}\{\frac{1}{2}[|b_1\rangle(|01\rangle+|10\rangle)\nonumber\\&
+(-1)^{b_2}\sqrt{2}|\overline{b_1}00\rangle]_{A_1B_1B_2} \otimes
\frac{1}{2} [|c_1\rangle(|01\rangle+|10\rangle)\nonumber\\&
+(-1)^{c_2}\sqrt{2}|\overline{c_1}00\rangle]_{A_2C_1C_2}\}\nonumber\\
= &
\frac{1}{4}QFT_{A_1A_2}[|b_1c_1\rangle\otimes(|01\rangle+|10\rangle)\otimes(|01\rangle+|10\rangle)\nonumber\\&
+
|b_1\overline{c_1}\rangle\otimes(-1)^{c_2}\sqrt{2}(|01\rangle+|10\rangle)\otimes|00\rangle
+ |\overline{b_1}c_1\rangle \nonumber\\&
\otimes(-1)^{b_2}\sqrt{2}|00\rangle\otimes(|01\rangle+|10\rangle) +
|\overline{b_1c_1}\rangle\nonumber\\&
\otimes(-1)^{b_2+c_2}2|00\rangle\otimes|00\rangle]_{A_1A_2B_1B_2C_1C_2}.
\end{align}
We notice that $QFT|xy\rangle =
\frac{1}{2}[|00\rangle+(-1)^{x}i^y|01\rangle\\+(-1)^{y}|10\rangle+(-1)^{x}(-i)^{y}|11\rangle]$,
 and thus
\begin{align}
|\psi(2)\rangle = &
\frac{1}{8}\{[|00\rangle+(-1)^{b_1}i^{c_1}|01\rangle+(-1)^{c_1}|10\rangle
+(-1)^{b_1}\nonumber\\&
(-i)^{c_1}|11\rangle]\otimes(|01\rangle+|10\rangle)\otimes(|01\rangle+|10\rangle)\nonumber\\&
+
[|00\rangle+(-1)^{b_1}i^{\overline{c_1}}|01\rangle-(-1)^{c_1}|10\rangle+(-1)^{b_1}\nonumber\\&
(-i)^{\overline{c_1}}|11\rangle]\otimes(-1)^{c_2}\sqrt{2}(|01\rangle+|10\rangle)\otimes|00\rangle
\nonumber\\
& +
[|00\rangle-(-1)^{b_1}i^{c_1}|01\rangle+(-1)^{c_1}|10\rangle-(-1)^{b_1}\nonumber\\&(-i)^{c_1}|11\rangle]\otimes(-1)^{b_2}\sqrt{2}|00\rangle\otimes(|01\rangle+|10\rangle)
\nonumber\\
& +
[|00\rangle-(-1)^{b_1}i^{\overline{c_1}}|01\rangle-(-1)^{c_1}|10\rangle-(-1)^{b_1}\nonumber\\&
(-i)^{\overline{c_1}}|11\rangle]\otimes(-1)^{b_2+c_2}2|00\rangle|00\rangle
\}_{A_1A_2B_1B_2C_1C_2}.
\end{align}The reduced density matrix in subsystem $A_1B_1B_2$ is
\begin{align}
\rho_{A_1B_1B_2} = &  _{A_2C_1C_2}\langle
000|\psi(2)\rangle\langle\psi(2)|000\rangle_{A_2C_1C_2} \nonumber\\
&+ _{A_2C_1C_2}\langle
100|\psi(2)\rangle\langle\psi(2)|100\rangle_{A_2C_1C_2}\nonumber\\
& + _{A_2C_1C_2}\langle
001|\psi(2)\rangle\langle\psi(2)|001\rangle_{A_2C_1C_2} \nonumber\\
&+ _{A_2C_1C_2}\langle
101|\psi(2)\rangle\langle\psi(2)|101\rangle_{A_2C_1C_2}\nonumber\\
& + _{A_2C_1C_2}\langle
010|\psi(2)\rangle\langle\psi(2)|010\rangle_{A_2C_1C_2} \nonumber\\
&+ _{A_2C_1C_2}\langle
110|\psi(2)\rangle\langle\psi(2)|110\rangle_{A_2C_1C_2}\nonumber\\
= & \frac{1}{8}(2|000\rangle\langle 000|+|001\rangle\langle
001|+|001\rangle\langle 010|\nonumber\\
&+|010\rangle\langle 001|+|010\rangle\langle
010|+2|100\rangle\langle 100|\nonumber\\
&+|101\rangle\langle 101|+|101\rangle\langle 110|+|110\rangle\langle
101|\nonumber\\
&+|110\rangle\langle 110|).
\end{align}The reduced density matrix in subsystem $A_2C_1C_2$ is
\begin{align}
\rho_{A_2C_1C_2} = &  _{A_1B_1B_2}\langle
000|\psi(2)\rangle\langle\psi(2)|000\rangle_{A_1B_1B_2}\nonumber\\
& + _{A_1B_1B_2}\langle
100|\psi(2)\rangle\langle\psi(2)|100\rangle_{A_1B_1B_2}\nonumber\\
& + _{A_1B_1B_2}\langle
001|\psi(2)\rangle\langle\psi(2)|001\rangle_{A_1B_1B_2}\nonumber\\
& + _{A_1B_1B_2}\langle
101|\psi(2)\rangle\langle\psi(2)|101\rangle_{A_1B_1B_2}\nonumber\\
& + _{A_1B_1B_2}\langle
010|\psi(2)\rangle\langle\psi(2)|010\rangle_{A_1B_1B_2}\nonumber\\
& + _{A_1B_1B_2}\langle
110|\psi(2)\rangle\langle\psi(2)|110\rangle_{A_1B_1B_2}\nonumber\\
= & \frac{1}{8}(2|000\rangle\langle 000|+|001\rangle\langle
001|+|001\rangle\langle 010|\nonumber\\
&+|010\rangle\langle 001|+|010\rangle\langle
010|+2|100\rangle\langle 100|\nonumber\\
&+|101\rangle\langle 101|+|101\rangle\langle 110|+|110\rangle\langle
101|\nonumber\\
&+|110\rangle\langle 110|).
\end{align}\qed

\subsection{\label{Locking Operator} Locking Operator}
We notice that the locking operator used in simultaneous
teleportation \cite{WYGL08} is not suitable for simultaneous dense
coding. To explain the reason, we calculate the reduced density
matrix in subsystem $A_1B$ when that locking operator is used,
instead of the quantum Fourier transform and Bell state being used as
the entanglement channel. The situations of using GHZ and W states
as entanglement channels are similar.

The locking operator used in simultaneous teleportation \cite{WYGL08} is
\begin{align}
U(LOCK)_{12}=H_1CNOT_{12}=\frac{1}{\sqrt{2}}\begin{bmatrix}1 & 0 & 0
& 1\\0 & 1 & 1 & 0\\1 & 0 & 0 & -1\\0 & 1 & -1 & 0\end{bmatrix}
,\end{align} where $H$ is the Hadamard transform, $CNOT$ is the
controlled-NOT gate, qubit 1 is the control qubit and qubit 2  the
target qubit. After step 1, the state of the composite system can be
written as
\begin{align}
|\psi'(1)\rangle =|\phi(b_1b_2)\rangle_{A_1B}\otimes
|\phi(c_1c_2)\rangle_{A_2C}.
\end{align}After step 2, the state of the composite system becomes
\begin{align}
|\psi'(2)\rangle = &
U(LOCK)_{A_1A_2}[\frac{1}{\sqrt{2}}(|0b_1\rangle+(-1)^{b_2}|1\overline{b_1}\rangle)_{A_1B}\nonumber\\
& \otimes \frac{1}{\sqrt{2}}
(|0c_1\rangle+(-1)^{c_2}|1\overline{c_1}\rangle)_{A_2C}]\nonumber\\
= &
\frac{1}{2}U(LOCK)_{A_1A_2}(|00b_1c_1\rangle+(-1)^{c_2}|01b_1\overline{c_1}\rangle
\nonumber\\
&
+(-1)^{b_2}|10\overline{b_1}c_1\rangle+(-1)^{b_2+c_2}|11\overline{b_1c_1}\rangle)_{A_1A_2BC}
\nonumber\\
= &
\frac{1}{2\sqrt{2}}[(|00\rangle+|10\rangle)|b_1c_1\rangle+(-1)^{c_2}(|01\rangle+|11\rangle)\nonumber\\
&|b_1\overline{c_1}\rangle+(-1)^{b_2}(|01\rangle-|11\rangle)|\overline{b_1}c_1\rangle+(-1)^{b_2+c_2}\nonumber\\
&(|00\rangle-|10\rangle)|\overline{b_1c_1}\rangle]_{A_1A_2BC}.
\end{align}The reduced density matrix in subsystem $A_1B$ is
\begin{align}
\rho'_{A_1B} = &  _{A_2C}\langle
0c_1|\psi'(2)\rangle\langle\psi'(2)|0c_1\rangle_{A_2C} +
_{A_2C}\langle
0\overline{c_1}|\psi'(2)\rangle\nonumber\\
&\langle\psi'(2)|0\overline{c_1}\rangle_{A_2C}+ _{A_2C}\langle
1c_1|\psi'(2)\rangle\langle\psi'(2)|1c_1\rangle_{A_2C}\nonumber\\
& + _{A_2C}\langle
1\overline{c_1}|\psi'(2)\rangle\langle\psi'(2)|1\overline{c_1}\rangle_{A_2C}\nonumber\\
= & \frac{1}{4}(|0b_1\rangle\langle 0b_1|+|0b_1\rangle\langle
1b_1|+|0\overline{b_1}\rangle\langle
0\overline{b_1}|\nonumber\\
&-|0\overline{b_1}\rangle\langle
1\overline{b_1}|+|1b_1\rangle\langle 0b_1|+|1b_1\rangle\langle
1b_1|\nonumber\\
&-|1\overline{b_1}\rangle\langle
0\overline{b_1}|+|1\overline{b_1}\rangle\langle 1\overline{b_1}|).
\end{align}Since $\rho'_{A_1B}$ is only dependent on $b_1$, we denote it as
$\rho'_{A_1B}(b_1)$. We have \begin{align}\rho'_{A_1B}(0) =
\frac{1}{4}
\begin{bmatrix}1 & 0 & 1 & 0\\0 & 1 & 0 & -1\\1 & 0 & 1 & 0\\0 & -1 & 0 &
1\end{bmatrix}\end{align} and \begin{align}\rho'_{A_1B}(1) =
\frac{1}{4}
\begin{bmatrix}1 & 0 & -1 & 0\\0 & 1 & 0 & 1\\-1 & 0 & 1 & 0\\0 & 1 & 0 &
1\end{bmatrix}.\end{align}

Since $\rho'_{A_1B}(0)\rho'_{A_1B}(1)=0$, Bob can distinguish these
two states and obtain $b_1$ by a POVM measurement on qubits $A_1B$.
Similarly, Charlie can also obtain $c_2$ by a POVM measurement on
qubits $A_2C$. Each receiver can learn 1 bit of his information
before they agree to simultaneously find out what Alice sends. The
aim of simultaneous dense coding is not achieved when $U(LOCK)$ is
used instead of the quantum Fourier transform.

\section{\label{Teleportation} Simultaneous Teleportation Using Quantum Fourier Transform}
In this section, we show that the quantum Fourier transform can
alternatively be used as the locking operator in simultaneous
teleportation. Let us begin with a brief review of simultaneous
teleportation between one sender and two receivers \cite{WYGL08}.
Suppose that Alice intends to teleport
$|\varphi_1\rangle_{T_1}=\alpha_1|0\rangle_{T_1}+\beta_1|1\rangle_{T_1}$
to Bob and
$|\varphi_2\rangle_{T_2}=\alpha_2|0\rangle_{T_2}+\beta_2|1\rangle_{T_2}$
to Charlie under the condition that Bob and Charlie must collaborate
to simultaneously obtain their respective quantum states. Initially,
Alice, Bob and Charlie share two EPR pairs
$\frac{1}{\sqrt{2}}(|00\rangle+|11\rangle)_{A_1B}$ and
$\frac{1}{\sqrt{2}}(|00\rangle+|11\rangle)_{A_2C}$, where qubits
$A_1A_2$ belong to Alice, qubits $B$ and $C$ belong to Bob and
Charlie respectively. Then the initial quantum state of the
composite system is
\begin{align}
|\chi(0)\rangle= &
|\varphi_1\rangle_{T_1}\otimes|\varphi_2\rangle_{T_2}\otimes
\frac{1}{\sqrt{2}}(|00\rangle+|11\rangle)_{A_1B}\nonumber\\
&\otimes \frac{1}{\sqrt{2}}(|00\rangle+|11\rangle)_{A_2C}.
\end{align}

The scheme of simultaneous teleportation consists of five steps.

(1) Alice performs the unitary transform $U(LOCK)$ on qubits
$A_1A_2$ to lock the entanglement channel. After that, the state of
the composite system becomes
\begin{align}
|\chi(1)\rangle = &
|\varphi_1\rangle_{T_1}\otimes|\varphi_2\rangle_{T_2}\otimes
U(LOCK)_{A_1A_2}[\frac{1}{\sqrt{2}}(|00\rangle\nonumber\\
&+|11\rangle)_{A_1B}\otimes
\frac{1}{\sqrt{2}}(|00\rangle+|11\rangle)_{A_2C}].
\end{align}

(2) Alice performs the Bell State Measurement on qubits $A_1T_1$ and
$A_2T_2$, like the original teleportation scheme \cite{BBCJPW93}. It
is easy to prove that $|\chi(1)\rangle$ can be written as
\begin{align}
|\chi(1)\rangle = & \frac{1}{4}
\sum_{x_1=0}^{1}\sum_{y_1=0}^{1}\sum_{x_2=0}^{1}\sum_{y_2=0}^{1}|\phi(x_1y_1)\rangle_{A_1T_1}|\phi(x_2y_2)\rangle_{A_2T_2}\nonumber\\
& U(LOCK)^{\dagger}_{BC}[U_B(x_1y_1)|\varphi_1\rangle_B\otimes
U_C(x_2y_2)|\varphi_2\rangle_C].
\end{align}

If the measurement results are $|\phi(x_1y_1)\rangle_{A_1T_1}$ and
$|\phi(x_2y_2)\rangle_{A_2T_2}$, the state of qubits $BC$ collapses
into
\begin{align}
|\chi(2)\rangle
=U(LOCK)_{BC}^{\dagger}[U_B(x_1y_1)|\varphi_1\rangle_B\otimes
U_C(x_2y_2)|\varphi_2\rangle_C].
\end{align}

(3) Alice sends the measurement results $(x_1,y_1)$ to Bob and
$(x_2,y_2)$ to Charlie.

(4) Bob and Charlie collaborate to perform $U(LOCK)$ on qubits $BC$, and then
the state of $BC$ becomes
\begin{align}
|\chi(3)\rangle= & U(LOCK)_{BC} U(LOCK)_{BC}^{\dagger}
[U_B(x_1y_1)|\varphi_1\rangle_B\nonumber\\
&\otimes
U_C(x_2y_2)|\varphi_2\rangle_C]\nonumber\\
= &U_B(x_1y_1)|\varphi_1\rangle_B\otimes
U_C(x_2y_2)|\varphi_2\rangle_C.
\end{align}

(5) Bob and Charlie perform $U(x_1y_1)$ and $U(x_2y_2)$ on qubits
$B$ and $C$ respectively to obtain $|\varphi_1\rangle$ and
$|\varphi_2\rangle$, respectively, like the original teleportation scheme
\cite{BBCJPW93}.

In the above simultaneous teleportation scheme, $U(LOCK)$ is used to
lock the entanglement channel. In Sec. \ref{Locking Operator}, we
have shown that $U(LOCK)$ is not suitable for simultaneous dense
coding, but, however, we find that the quantum Fourier transform can
alternatively be used as the locking operator in simultaneous
teleportation.

Let us suppose that Alice is the sender, Bob$_i (1\leqslant
i\leqslant N)$ are the receivers. Alice intends to send the unknown
quantum states
$|\varphi_i\rangle_{T_i}=(\alpha_i|0\rangle+\beta_i|1\rangle)_{T_i}$
to Bob$_i$ under the condition that all the receivers must
collaborate to simultaneously obtain
$(\alpha_i|0\rangle+\beta_i|1\rangle)_{T_i}$. Initially, Alice and
each receiver share an EPR pair $\frac{1}{\sqrt{2}}(|00\rangle
+|11\rangle)_{A_iB_i}$. The initial quantum state of the composite
system is
\begin{align}
|\chi'(0)\rangle & = \frac{1}{\sqrt{2^N}} \bigotimes_{i=1}^N
|\varphi_i\rangle_{T_i}\bigotimes_{i=1}^N
(|00\rangle + |11\rangle)_{A_iB_i} \nonumber\\
& = \frac{1}{\sqrt{2^N}} \bigotimes_{i=1}^N
|\varphi_i\rangle_{T_i}\sum_{m=0}^{2^N-1} |m\rangle_{A_1\dots A_N}
|m\rangle_{B_1\dots B_N}.
\end{align}

The scheme of simultaneous teleportation consists of five steps.

(1) Alice performs the quantum Fourier transform
$|j\rangle\rightarrow\frac{1}{\sqrt{2^N}}\sum_{k=0}^{2^N-1}e^{2\pi
ijk/2^N}|k\rangle$ on qubits $A_1\dots A_N$ to lock the entanglement
channel. After that, the state of the composite system becomes
\begin{align}
|\chi'(1)\rangle & = QFT_{A_1\dots
A_N}|\chi'(0)\rangle\nonumber\\
& = \frac{1}{2^N}\bigotimes_{i=1}^N |\varphi_i\rangle_{T_i}
\sum_{m=0}^{2^N-1}\sum_{k=0}^{2^N-1} \omega^{mk}|k\rangle_{A_1\dots
A_N}|m\rangle_{B_1\dots B_N} \nonumber\\
& =\frac{1}{2^N} \sum_{k=0}^{2^N-1}\sum_{m=0}^{2^N-1}\omega^{mk}
\bigotimes_{i=1}^N(|k_i\rangle_{A_i}|\varphi_i\rangle_{T_i})|m\rangle_{B_1\dots
B_N}
 ,
\end{align}
where $k_i$ is the $i$th bit of $k$, $\omega = e^{2\pi i/2^N}$.

(2) Alice performs the Bell State Measurement on each pair of
$A_iT_i$.

We have
\begin{align}
\bigotimes_{i=1}^{N} I_{A_iT_i} = & \bigotimes_{i=1}^{N}
\sum_{x_i=0}^{1} \sum_{y_i=0}^{1} |\phi(x_iy_i)\rangle_{A_iT_i}\
_{A_iT_i}\langle\phi(x_iy_i)| \nonumber\\
= & \sum_{x_1=0}^{1} \sum_{y_1=0}^{1} \dots \sum_{x_N=0}^{1}
\sum_{y_N=0}^{1} \bigotimes_{i=1}^{N} |\phi(x_iy_i)\rangle_{A_iT_i}\nonumber\\
&
_{A_iT_i}\langle\phi(x_iy_i)| \nonumber\\
= & \sum_{x_1=0}^{1} \sum_{y_1=0}^{1} \dots \sum_{x_N=0}^{1}
\sum_{y_N=0}^{1} \bigotimes_{i=1}^{N} |\phi(x_iy_i)\rangle_{A_iT_i}\nonumber\\
& \bigotimes_{i=1}^{N}\ _{A_iT_i}\langle\phi(x_iy_i)|
\end{align}

and
\begin{align}
& \bigotimes_{i=1}^{N}\
_{A_iT_i}\langle\phi(x_iy_i)|\chi'(1)\rangle\nonumber\\ = &
\frac{1}{\sqrt{2^N}} \sum_{k=0}^{2^N-1} \bigotimes_{i=1}^N\
_{A_iT_i} (\langle0x_i|+(-1)^{y_i}\langle1\overline{x_i}|)
(\alpha_i|k_i0\rangle\nonumber\\
&+ \beta_i |k_i1\rangle)_{A_iT_i} \frac{1}{\sqrt{2^N}}
\sum_{m=0}^{2^N-1}\omega^{mk}
|m\rangle_{B_1\dots B_N}\nonumber\\
= & \frac{1}{\sqrt{2^N}} \sum_{k=0}^{2^N-1} \prod_{i=1}^N
[\delta_{k_i0}(\delta_{x_i0}\alpha_i+\delta_{x_i1}\beta_i)+\delta_{k_i1}(-1)^{y_i}\nonumber\\
&(\delta_{x_i1}\alpha_i+\delta_{x_i0}\beta_i)]
QFT_{B_1\dots B_N} |k\rangle_{B_1\dots B_N}\nonumber\\
= & \frac{1}{\sqrt{2^N}} QFT_{B_1\dots B_N} \bigotimes_{i=1}^N
[(\delta_{x_i0}\alpha_i+\delta_{x_i1}\beta_i)|0\rangle
\nonumber\\
&+(-1)^{y_i}(\delta_{x_i0}\beta_i + \delta_{x_i1}\alpha_i)|1\rangle]_{B_i}\nonumber\\
= & \frac{1}{\sqrt{2^N}} QFT_{B_1\dots B_N} \bigotimes_{i=1}^N
U(x_iy_i) (\alpha_i|0\rangle +\beta_i|1\rangle)_{B_i}.
\end{align}

Thus, $|\chi'(1)\rangle$ can be written as
\begin{align}
|\chi'(1)\rangle = & \bigotimes_{i=1}^{N} I_{A_iT_i}
|\chi'(1)\rangle
\nonumber\\
= & \sum_{x_1=0}^{1} \sum_{y_1=0}^{1} \dots \sum_{x_N=0}^{1}
\sum_{y_N=0}^{1} \bigotimes_{i=1}^{N} |\phi(x_iy_i)\rangle_{A_iT_i}\nonumber\\
& \bigotimes_{i=1}^{N}\ _{A_iT_i} \langle\phi(x_iy_i)
|\chi'(1)\rangle\nonumber\\
= & \frac{1}{\sqrt{2^N}} \sum_{x_1=0}^{1} \sum_{y_1=0}^{1} \dots
\sum_{x_N=0}^{1} \sum_{y_N=0}^{1} \bigotimes_{i=1}^{N}
|\phi(x_iy_i)\rangle_{A_iT_i} \nonumber\\
&QFT_{B_1\dots B_N} \bigotimes_{i=1}^{N}\
U(x_iy_i)|\varphi_i\rangle_{B_i}.
\end{align}

If the measurement result of qubits $A_iT_i$ is
$|\phi(x_iy_i)\rangle$, the state of qubits $B_1\dots B_N$ collapses
into
\begin{align}
|\chi'(2)\rangle = QFT_{B_1\dots B_N} \bigotimes_{i=1}^{N}\
U(x_iy_i)|\varphi_i\rangle_{B_i}.
\end{align}

(3) Alice sends the measurement result $(x_i, y_i)$ to each Bob$_i$.

(4) All the receivers collaborate to perform $QFT^\dagger$ on qubits
$B_1\dots B_N$, the state of $B_1\dots B_N$ becomes
\begin{align}
|\chi'(3)\rangle = & QFT^\dagger_{B_1\dots B_N}QFT_{B_1\dots B_N}
\bigotimes_{i=1}^{N}\ U(x_iy_i)|\varphi_i\rangle_{B_i} \nonumber\\
= & \bigotimes_{i=1}^{N}\ U(x_iy_i)|\varphi_i\rangle_{B_i}.
\end{align}

(5) Each Bob$_i$ performs $U(x_iy_i)$ on qubit $B_i$ to obtain
$|\varphi_i\rangle$.

\section{\label{Conclusion} Conclusion}
In summary, we have proposed a simultaneous dense coding scheme
between one sender and two receivers, the aim of which is for the
receivers to simultaneously obtain their respective messages. This
scheme may be used in a security scenario. For example, Alice wants
Bob and Charlie to simultaneously carry out two confidential
commercial activities under the condition that the sensitive
information of each activity is only revealed to who is in charge of
that activity. We have also shown that the quantum Fourier
transform, which has been implemented using cavity quantum
electrodynamics (QED) \cite{SZ02}, nuclear magnetic resonance (NMR)
\cite{VSBYCC00,FLXZ00,WPFLC01,VSBYSC01,DS05} and coupled
semiconductor double quantum dot (DQD) molecules \cite{DYC08}, can
act as the locking operator both in simultaneous dense coding and
teleportation.

This work is supported  by the National Natural Science Foundation
(Nos. 60573006, 60873055), and the Research Foundation for the
Doctorial Program of Higher School of Ministry of Education (No.
20050558015), and NCET of China.

\end{document}